\documentclass[aps,prd,twocolumn,groupedaddress]{revtex4-1}

\usepackage[dvips]{graphicx,color}
\usepackage{ amssymb }

\def\nat{Nature\ }

\def\aap{Astron.\ Astrophys.\ }
\def\araa{Ann.\ Rev.\ Astron.\ Astrophys.\ }
\def\apj{Astrophys.\ J.\ }
\def\apjl{Astrophys.\ J.\ Lett.\ }
\def\apjs{Astrophys.\ J.\ Suppl.\ }
\def\aj{Astron.\ J.\ }
\def\mnras{Mon.\ Not.\ Roy.\ Astron.\ Soc.\ }
\def\physrep{Phys.\ Rept.\ }
\def\prd{Phys.\ Rev.\ D\ }
\def\apss{Astrophys.\ Space\ Sci.\ }
\def\jcap{JCAP}

\begin{document}
\title{Stellar convective cores as dark matter probes}
\author{Jordi Casanellas$^1$, Isa Brand\~ao$^2$, Yveline Lebreton$^{3,4}$}
\affiliation{$^1$Max Planck Institut f\"ur Gravitationsphysik (Albert-Einstein-Institut)\\
$^2$Centro de Astrof\'isica and Faculdade de Ci\^encias, Universidade do Porto\\
$^3$Observatoire de Paris, GEPI, CNRS UMR 8111, F-92195 Meudon, France\\
$^4$Institut de Physique de Rennes, Universit\'e de Rennes 1, CNRS UMR 6251, F-35042 Rennes, France}
\email{jordi.casanellas@aei.mpg.de}
\begin{abstract}
The recent detection of a convective core in a main-sequence solar-type star is used here to test particular models of dark matter (DM) particles, those with masses and scattering cross sections in the range of interest for the DM interpretation of the positive results in several DM direct detection experiments. If DM particles do not effectively self-annihilate after accumulating inside low-mass stars (\textit{e.g.} in the asymmetric DM scenario) their conduction provides an efficient mechanism of energy transport in the stellar core. For main-sequence stars with masses between 1.1 and 1.3~M$_{\odot}$, this mechanism may lead to the suppression of the inner convective region expected to be present in standard stellar evolution theory. The asteroseismic analysis of the acoustic oscillations of a star can prove the presence/absence of such a convective core, as it was demonstrated for the first time with the \textit{Kepler} field main-sequence solar-like pulsator, KIC 2009505. Studying this star we found that the asymmetric DM interpretation of the results in the CoGeNT experiment is incompatible with the confirmed presence of a small convective core in KIC 2009505.
\end{abstract}
\maketitle

\section{Introduction}
The quest to unveil the constituents of the Dark Matter (DM) of the Universe is nowadays one of the most challenging goals in modern physics~\cite{2014Natur.507...29L}. Several models beyond the standard model have been put forward to explain the properties of DM in terms of new, still undetected particles~\cite{1985NuPhB.253..375S,2010ARA&A..48..495F}. Colliders, ground-based telescopes, satellites, and underground experiments are trying to detect the hypothetical non-gravitational interactions of these DM particles aiming at the discovery of their elusive nature~\cite{2005PhR...405..279B,2012PhRvD..85e6011F,2014PhRvD..89d2001A,2015arXiv150303379S}. Interestingly, several direct detection experiments have reported signals which could have arisen from collisions of low-mass DM particles with the nuclei in the detectors~\cite{Bernabei:2010mq,2011PhRvL.107n1301A,2012EPJC...72.1971A,Agnese:2013rvf}. However, this interpretation is challenged by the null results in many other experiments~\cite{2011PhRvL.107e1301A,Aprile:2013doa,Archambault:2012pm,Felizardo:2011uw,Akerib:2013tjd,2015arXiv150300008A}. Whether these opposed results are fully incompatible or not is still a matter of debate~\cite{Hooper:2013cwa,2013PhRvD..88e6003B,Davis:2014bla}.\\

Here we apply a novel approach to test the existence of DM particles with properties as those that would explain the positive results of direct detection experiments by looking for the particular effects that they would produce in stars slightly more massive than the Sun. As it will be shown, this method offers an inexpensive, complementary probe on the plausibility of some DM candidates. The models of DM that can be tested with our approach are non-(or feebly) annihilating particles, with the low masses and relatively large scattering cross-sections with baryons required to explain the results of the DAMA~\cite{Bernabei:2010mq}, CoGeNT~\cite{2011PhRvL.107n1301A} and CDMS-Si~\cite{Agnese:2013rvf} experiments. 
Theories that predict these low-mass Weakly Interacting Massive Particles (WIMPs) include Asymmetric DM (ADM, \cite{2013IJMPA..2830028P,2014AIPC.1604..389K,2014PhR...537...91Z}), although our analysis applies to any model of DM for which the number of DM particles accumulated in the core of a star is not significantly depleted due to annihilations or decays, independently of the mechanism behind this property. In the context of ADM, although the self-annihilation cross-section of the DM particles may be large, their annihilation inside stars is prevented by the lack of antiparticles, originated from an asymmetry in the particle/antiparticle abundances related to the baryonic one.\\

The properties of ADM have already been constrained using observations of other type of stars. Neutron stars, being very dense, could capture and accumulate huge quantities of DM particles~\cite{2014JCAP...05..013G,2014PhRvD..90h3507C}, leading to changes in the properties of the star \cite{2012JCAP...10..031L,2013PhRvD..87l3506L,2014PhRvC..89b5803X} and the hypothetical creation of a black hole that could destroy it \cite{2008PhRvD..77d3515B,2011PhRvL.107i1301K,2012PhRvL.108s1301K,2013PhRvD..87l3507B,2013PhRvD..87l3537K,2014PhRvL.113s1301B,2014PhRvD..89a5010B}. Although the constraints on ADM from this approach are very stringent, they have the uncertainty associated with the complexity of the high-energy physics involved in the self-collapse of DM in the core of a neutron star~\cite{Bertoni:2013bsa,2014PhRvD..90d3512K}.\\

\begin{table*}[!t]
\centering
 \begin{tabular}{l c c c c c c}
 \hline\hline
  & $T_\mathrm{eff}$ & $\log (g)$ & $[Z/X]_s$ & $R_\mathrm{cc}$ & $\langle \Delta\nu \rangle _{012}$ & $|S\{ \Delta\nu_{0} \, r_{010}\}|$\\
  & ( K ) & & & ( R$_{\star}$ ) & ( $\mu$Hz ) & \\
 \hline
  - Observations: &&&&&& \\
  \vspace{2mm}
  \textit{Dushera} & 6200 $\pm$ 200 & 4.30 $\pm$ 0.2 & 0.019 $\pm$ 0.006\footnote{the observed [Fe/H]$=\log(Z/X) - \log(Z/X)_{\odot}$ was converted to the surface $[Z/X]$ assuming equal iron and metal abundances.
  } & 0.061-0.071\footnote{the range in the convective core radius is that of the best fit models in Ref.~\cite{2013ApJ...769..141S}} & 88 $\pm$ 0.6 & 0.0032 $\pm$ 0.0006 \\
  - Models: &&&&&& \\
  Standard (no DM) & 6222 $\pm$ 107 & 4.21 $\pm$ 0.01 & 0.019 $\pm$ 0.004 & 0.069 $\pm$ 0.006 & 87.3 $\pm$ 1.1 & 0.0028 $\pm$ 0.0011 \\
  DM model 1 & 6208 $\pm$ 110 & 4.21 $\pm$ 0.01 & 0.019 $\pm$ 0.004 & 0.069 $\pm$ 0.006 & 87.3 $\pm$ 0.9 & 0.0028 $\pm$ 0.0012 \\
  DM model 2 & 6208 $\pm$ 111 & 4.21 $\pm$ 0.01 & 0.019 $\pm$ 0.004 & 0.069 $\pm$ 0.006 & 87.2 $\pm$ 1.1 & 0.0028 $\pm$ 0.0012 \\
  DM model 3 & 6255 $\pm$ 103 & 4.21 $\pm$ 0.01 & 0.020 $\pm$ 0.004 & 0.001 $\pm$ 0.008 & 87.1 $\pm$ 1.0 & 0.0006 $\pm$ 0.0004 \\
  DM model 4 & 6219 $\pm$ 107 & 4.21 $\pm$ 0.01 & 0.020 $\pm$ 0.004 & 0.028 $\pm$ 0.016 & 87.3 $\pm$ 0.9 & 0.0028 $\pm$ 0.0015 \\
  \hline\hline
 \end{tabular}\\
   \caption{Parameters determined from ground-based high-resolution spectroscopy~\cite{2012MNRAS.423..122B} and \textit{Kepler} high-precision photometry~\cite{2011AJ....142..112B} for \textit{Dushera}, and distribution of the same parameters in the models obtained from our grids.\label{tab-params}}
\end{table*}
On the other hand, most of the processes governing the interior of main-sequence stars are well understood and tested by precise observations, including the solar neutrinos and helioseismology~\cite{2011RPPh...74h6901T}. In the Sun, the accumulated ADM particles would quickly thermalize in a small region of the solar core and their conduction would modify the solar central temperature and density, leaving potentially strong imprints in the solar neutrino fluxes and oscillations \cite{Lopes:2002gp,2010PhRvL.105a1301F,2010PhRvD..82h3509T,2010Sci...330..462L}. Remarkably, stars with masses slightly greater than the Sun can have stronger structural changes in their interior due to ADM. It has been shown that if DM does not self-annihilate after accumulating inside stars with masses between 1.1 and $1.3\;\textmd{M}_{\odot}$, their conduction may remove energy from the stellar nucleus efficiently enough to suppress the small convective core that these stars are expected to develop in the main sequence~\cite{2013ApJ...765L..21C}.\\

In the last years, the space-based asteroseismic missions \textit{CoRoT} and \textit{Kepler} have detected solar-like oscillations in thousands of stars~\cite{2013ARA&A..51..353C}. The high precision in the determination of the oscillation frequencies of these stars has allowed not only measurements of their masses and radius with unprecedented accuracy, but also opened a window into the stellar interiors, including the discrimination between He-burning and inert cores in red giants~\cite{2011Natur.471..608B}. In this work, we analyse KIC 12009504, the first \textit{Kepler} main-sequence solar-like pulsator to have its convective core detected thanks to asteroseismic studies performed on this star~\cite{2013ApJ...769..141S}. We use this detection to rule out DM models that would prevent the formation of such a convective region.\\

This article is organized as follows. The modeling of the star and the DM particles are described in Section~\ref{sec-modeling}. Section~\ref{sec-dmimpact} reviews the impact of DM conduction in stars. In Section~\ref{sec-astrocc} we present the diagnostic tools used to infer the presence/absence of a convective core from the stellar acoustic oscillations. In Section~\ref{sec-results} we show the results of our simulations and how they compare with the observations of KIC 12009504, and finally in Section~\ref{sec-concl} we summarize and discuss the conclusions of this work.\\

\section{Modeling}
\label{sec-modeling}

\subsection{Stellar modeling}
\label{ssec-stelmod}
KIC 12009504, also known as \textit{Dushera}, is a main-sequence solar-like pulsator which has been observed by ground-based high-resolution spectroscopy~\cite{2012MNRAS.423..122B} and by the high-precision photometry of the \textit{Kepler} mission ~\cite{2011AJ....142..112B}. These observations led to measurements of its effective temperature $T_\mathrm{eff}$, surface gravity $\log (g)$ and surface metallicity $[Z/X]_s$ which are shown in Table~\ref{tab-params}. For the modeling of \textit{Dushera} we assumed the large uncertainties in $T_\mathrm{eff}$ quoted in Table~\ref{tab-params}, following the approach in a previous thorough analysis of this star \cite{2013ApJ...769..141S}, due to an inconsistency between different spectroscopic analysis \cite{2008A&A...487..373S}.\\

Solar-like oscillations have been observed in \textit{Dushera}, with the individual frequencies of 34 modes precisely determined in Ref.~\cite{2012A&A...543A..54A}, allowing an accurate estimation of its properties. In particular, its mass has been estimated with a very small uncertainty ($\pm0.03 \;$M$_{\odot}$) through the fitting of individual frequencies on the basis of stellar models~\cite{2014ApJS..214...27M}. However, for our model fitting we adopted a conservative approach and considered the broader mass range obtained by Ref.~\cite{2013ApJ...769..141S}: $1.23 \pm 0.12$M$_\odot$. The latter uncertainty comes from a combined analysis of the modeling of \textit{Dushera} by several teams, using different stellar evolution codes, pulsation codes, and fitting techniques, including the fitting to frequency ratios which are sensitive to the stellar interior.\\

The modeling of \textit{Dushera} was performed using a modified version of CESAM stellar evolution code~\cite{2008Ap&SS.316...61M}, including microscopic diffusion~\cite{1993ASPC...40..246M}, convection through the mixing-length theory~\cite{1958ZA.....46..108B} with overshooting of the convective core implemented as $d_{ov}=\alpha_{ov}\min(R_\mathrm{cc}, H_{p})$, where $d_{ov}$ is the overshooting distance, $R_\mathrm{cc}$ is the radius of the Schwarzschild convective core, $H_{p}$ the pressure scale height and $\alpha_{ov}$ a free parameter~\cite{1997ASSL..225...23R}, and stellar heavy element mixture as the solar one~\cite{2009ARA&A..47..481A}.\\

An extensive grid of stellar models was computed, with stellar masses in the range $M_{\star}=1.1 - 1.35\,\textmd{M}_{\odot}$, initial metallicities between $Z=0.012 - 0.024$, initial helium abundances between $Y_0=0.26 - 0.30$, overshooting with efficiency parameter $\alpha_{ov}=0.05, 0.15$, and convection with $\alpha_{MLT}=$1.6, 1.8 and 2.0. The same grid was computed for the standard stellar evolution scenario, without DM, and also considering the impact of the different models of DM particles described in Table~\ref{tab-DM} and in Section~\ref{ssec-DMmod}.\\

From all the stellar models computed in our grids, we selected those that reproduce the observed properties of \textit{Dushera} ($T_\mathrm{eff}$, $\log (g)$, $[Z/X]_s$) within 1-$\sigma$ and the large frequency separation, $\Delta\nu$, within 2-$\sigma$. We allowed for a larger error in $\Delta\nu$ because, at this stage of the modeling pipeline, this parameter was estimated using scaling relations based on an asymptotic approximation: $\Delta\nu = \Delta\nu_{\odot} \left(M/M_{\odot}\right)^{0.5} \left(T_\mathrm{eff}/T_{eff,\odot}\right)^{3} \left(L/L_{\odot}\right)^{-0.75}$ (see \cite{2013A&A...550A.126M,2013MNRAS.434.1668H}). Our grid-based selection procedure is the standard to determine stellar properties from their oscillations (see \textit{e.g.} Ref.~\cite{2014ApJS..210....1C}) and is comparable to 7 of the 8 methods compiled in Ref.~\cite{2013ApJ...769..141S} to simulate the same star. An alternative would be to use Markov Chain Monte Carlo or genetic algorithms, which can be more precise in providing the best-fit model due to the finite resolution of grid-based methods~\cite{2012MNRAS.427.1847B,2014ApJS..214...27M}.\\

In addition, we estimate that the fraction of valid models of \textit{Dushera} that were discarded due to the limits imposed in our grids is below 5\% (the mass and metallicities of the selected models have mean and standard deviations of 1.19$\pm$0.05 M$_{\odot}$ and 0.018$\pm$0.003, respectively).\\

This procedure resulted in more than 30,000 valid models of \textit{Dushera}. For all these models, the acoustic oscillation frequencies were computed using the ADIPLS code~\cite{art-Ch-Dals2008Ap&SS}. With these frequencies we recalculated the large separation $\Delta\nu$ and computed the seismic parameters described in Section~\ref{sec-astrocc}.\\
\subsection{Dark Matter modeling}
\label{ssec-DMmod}
The present status of direct dark matter detection experiments is characterized by the tension between the promising positive results in some experiments~\cite{Bernabei:2010mq,2011PhRvL.107n1301A,2012EPJC...72.1971A,Agnese:2013rvf} and the robust incompatible limits set by the null results in others~\cite{2011PhRvL.107e1301A,Aprile:2013doa,Archambault:2012pm,Felizardo:2011uw,Akerib:2013tjd,2015arXiv150300008A}. While some models try to simultaneously explain all the results, other analysis suggest that backgrounds or biased analysis may explain the controversy~\cite{Hooper:2013cwa,Davis:2014bla}. In this context, the method we propose here provides an alternative test of the existence of the low-mass WIMPs with ``large'' scattering cross-section on nucleons that best fit the signals in some of these experiments.\\
\begin{table}[!t]
\centering
 \begin{tabular}{l r c r}
 \hline\hline
  & $m_{\chi}$ & $\sigma_{\chi}$ & Experiment \\
  & ( GeV ) & ( cm$^2$ ) & \\ 
 \hline
 DM model 1 & 7 & $10^{-40}$ & \footnotesize{DAMA+CDMS+CoGeNT}\footnote{interpreted in terms of SI WIMP-nucleon interactions~\cite{2014PDU.....5....1A}}\\
 DM model 2 & 13 & $10^{-36}$ & \footnotesize{DAMA}\footnote{interpreted in terms of SD WIMP-proton interactions~\cite{2013PhRvD..88e6003B}} \\
 DM model 3 & 8 & $10^{-33}$ & \footnotesize{CoGeNT}$^\textmd{\scriptsize{b}}$ \\
 DM model 4 & 10 & $10^{-32}$ & \footnotesize{CDMS-Si}$^\textmd{\scriptsize{b}}$ \\ 
 \hline\hline
 \end{tabular}
  \caption{Characteristics of the models of dark matter particles tested in this work.\label{tab-DM}}
\end{table}

The properties of the models of DM tested here, shown in Table~\ref{tab-DM}, are those that can explain the signals in DAMA, CDMS-Si and CoGeNT in terms of spin-dependent (SD) WIMP interactions on protons~\cite{2013PhRvD..88e6003B} (labeled ``DM model 2, 3 and 4'', respectively), and the model that best fits all these experiments in terms of spin-independent (SI) interactions on nucleons~\cite{2014PDU.....5....1A} (labeled ``DM model 1''). The DM scattering off nuclei through SD (axial-vector) and SI (scalar) interactions is implemented in our models separately, and we tested models with either pure SD interactions (DM models 2, 3 and 4) or pure SI interactions (DM model 1). However, more realistic models may interact simultaneously through the two classes of couplings: SD interactions with hydrogen nuclei and SI interactions with all the elements of the stellar plasma. As stated before, we stress that our results apply in general to any model that predicts DM particles with masses and scattering cross sections similar to those listed in Table~\ref{tab-DM}, as long as the number of DM particles concentrated inside stars is not notably reduced by mechanisms such as annihilations or decays.\\ 

We have analyzed the standard scenario of elastic scattering between DM particles and nucleons. For the sake of simplicity, here we have not considered other scenarios with more complex interactions, such as long-range interactions, velocity or momentum-dependencies, isospin violation, exothermic or inelastic scatterings, etc., although they may boost the DM impact on stars~\cite{2012ApJ...752..129L,2014ApJ...795..162L,2014JCAP...04..019V,2015arXiv150404378V}. In principle, the method of DM search proposed here could be applied to constrain the parameters describing the WIMP-nucleus interactions in more general frameworks such as DM effective field theory~\cite{2013JCAP...02..004F,2015arXiv150303379S} or Minimal DM~\cite{2001NuPhB.619..709B,2006NuPhB.753..178C,2015arXiv150301119B}.\\

The inclusion of the impact of DM on the stellar properties was computed following the prescriptions of Gould~\cite{art-Gould1987} for the capture rate (using modified subroutines of the {\sf DarkSUSY} code~\cite{art-GondoloEdsjoDarkSusy2004}), and Gould and Raffelt~\cite{Gould:1989ez} for the energy transport by DM conduction. The reader is addressed to the references above for a thorough description of the formalisms used to model the impact of DM on stars. Nonetheless, it is worth highlighting some simple relations that illustrate the main dependencies of the processes involved.\\

The efficiency of the capture of DM particles by a star is proportional to the DM-nucleon scattering cross-section and the density of DM around the star, $C_{\chi} \propto \sigma_{\chi} \rho_{\chi}$, and also depends on the velocity distribution of the DM particles. We assumed a canonical DM halo model with a Maxwell-Boltzmann velocity distribution, a dispersion $\bar{v_{\chi}}=270\;$km s$^{-1}$ and a local DM density around the star $\rho_{\chi}=0.4\;$GeV$\;$cm$^{-3}$, as estimated in the solar neighborhood~\cite{Catena:2009mf}. Departures from the standard Maxwellian velocity distribution are well motivated by simulations and observations \citep{art-Navarroetal2010MNRAS,art-Merrittetal2006AJ} and have been extensively explored to assess the event rate in DM direct detection experiments~\cite{2009MNRAS.395..797V,2014PhRvD..89f3513M}. We address the reader to Ref.~\cite{Lopes:2011rx}, where the impact in the capture rate $C_{\chi}$ of the uncertainties in $\rho_{\chi}$ and $v_{\chi}$ was studied.\\

The captured DM particles quickly thermalize with the stellar plasma and concentrate in the core of the star with a distribution characterized by a radius that shrinks with the DM mass: $r_{\chi} \propto m_{\chi}^{-0.5}$. Consequently, DM particles with low masses can influence a broad region 
transporting energy by conduction ($r_{\chi}\approx5\%\;\textmd{R}_{\star}$ for $m_{\chi}=10\;$GeV), whereas DM particles with high masses distribute too concentratedly to decisively impact the star by this mechanism.\\ 

WIMPs remove energy from the stellar core and bring it to a distance characterized by their mean free path inside the star $l_{\chi}(r)$. This parameter defines, together with $r_{\chi}$, the Knudsen number of the system $K=l_{\chi}(0)/r_{\chi}$, which shows whether WIMPs transport energy locally or not. The luminosity transported by DM is:
\begin{eqnarray}
L_\mathrm{\chi,trans}(r) &=& 4 \pi r^2 \, n_\mathrm{\chi,LTE}(r) \, l_{\chi} \, \kappa(r) \nonumber\\
&& \times \left(\frac{k_B T(r)}{m_{\chi}}\right)^{1/2} k_B \, \frac{dT}{dr} \, f(K) \, h(r),
\end{eqnarray}
where $\kappa(r)$ is the thermal conductivity for a gas of WIMPs and the elements of the stellar plasma, found by interpolating in the table calculated numerically by Ref.~\cite{1990ApJ...352..654G}. $h(r)$ and $f(K)$ are suppression factors, the latter introduced to extend the expression for the energy transport in local thermal equilibrium (LTE) to regimes with a larger WIMP mean free path within the stellar plasma. The real WIMP distribution inside the star, $n_{\chi}(r)$, is found by interpolating between the LTE and the isothermal distributions (see Ref.\cite{Scott:2008ns} for the details).\\

\section{Dark Matter impact on stellar properties}
\label{sec-dmimpact}
The additional mechanism of stellar energy transport by DM conduction described above was implemented in the stellar evolution code CESAM. Our modified version of the code has been tested in the case of the Sun and other stars \cite{Casanellas:2009dp,2012ApJ...757..130L,2013ApJ...765L..21C} producing results in agreement with those obtained independently by other groups using different codes~\cite{Scott:2008ns,2010PhRvD..82j3503C}.\\

The main impact of DM cooling in the properties of main-sequence stars occurs in their core, and consists in a reduction in the central temperature of the star and an increase in their central density (see Ref.~\cite{2013ApJ...765L..21C}). Strong deviations in the global properties of the star, such as its luminosity and its effective temperature, are only expected if the star is located in environments with DM densities orders of magnitude higher than those considered here ~\cite{2012PhRvL.108f1301I}.\\

\begin{figure}[!t]
\centering
\includegraphics[scale=0.9]{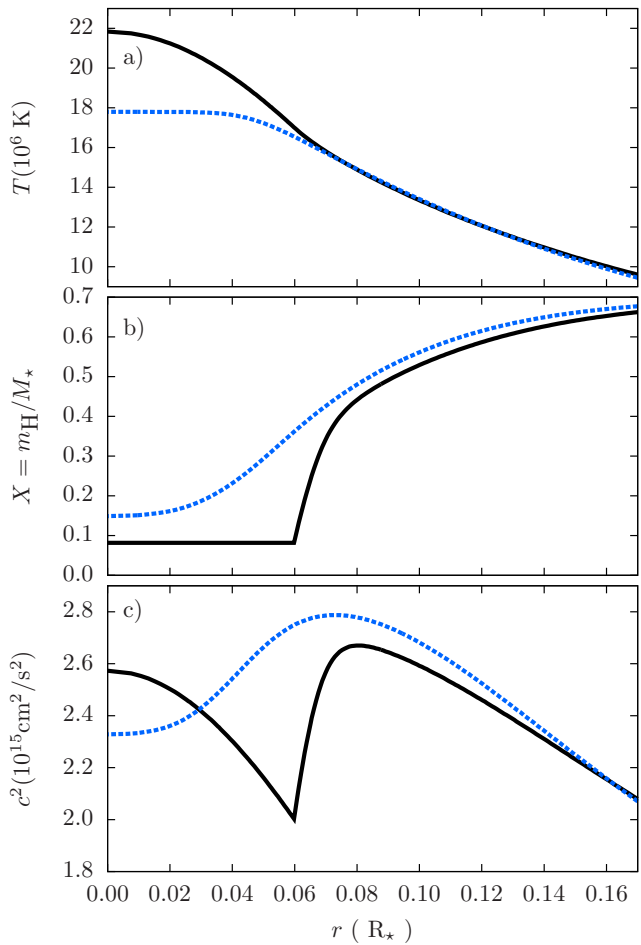}
\caption{a) Temperature, b) hydrogen abundance in mass fraction, and c) adiabatic sound-speed profiles for representative models of \textit{Dushera} with a convective core at $R_\mathrm{cc}=0.07\;$R$_{\star}$ (black continuous line, corresponding to a model not influenced by DM) and without it (blue dashed line, corresponding to a model strongly impacted by DM conduction).}
\label{fig-multirad}
\end{figure}

In this work we are especially interested in the changes on the temperature profile in the core of the star. For a star like \textit{Dushera}, the DM conduction will cool the central $\sim$5\% of the star, leading to an isothermal core which is only present when the impact of DM is included (see Figure~\ref{fig-multirad}.a)). As it was shown in Ref.~\cite{2013ApJ...765L..21C}, this cooling may lead to a strong structural change in the core of the star, suppressing convection in the central region. Convective instabilities arise when the temperature gradient is so steep that rising bubbles of stellar plasma do not cool rapidly enough through adiabatic expansion and they continue to rise. Due to the large energy flux resulting from the CNO hydrogen burning cycle, main-sequence stars with masses slightly greater than the Sun are known to develop a small convective core in the standard scenario of stellar evolution. In contrast, this convective core would be suppressed if DM particles as those studied here contribute to cool the stellar core.\\

Within the context of the ADM models considered here, the DM particles accumulate inside the star during its evolution without annihilating. This is a crucial factor in order to have enough DM particles to noticeably cool the stellar core due to DM conduction. Depending on the unknown DM properties, mainly $m_{\chi}$ and $\sigma_{\chi}$, the reduction of the temperature gradient in the nucleus can be efficient enough to prevent the convective instability in the core of \textit{Dushera}.\\

\section{Asteroseismic signature of convective cores}
\label{sec-astrocc}
The presence of a convective core in the center of a star has a strong impact on the propagation of acoustic waves. The mixing in the stellar plasma induced by convection homogenizes the chemical composition in the convective regions, thus creating a discontinuity in the border with the radiative regions. This is depicted in Figure~\ref{fig-multirad} b), where the radial profiles of the H abundance, $X(r)$, of two models of \textit{Dushera} are plotted. An abrupt change in the H abundance can be seen in the DM-free model, with a convective core at $R_ {cc}\simeq0.07\;\textmd{R}_{\star}$, whereas the changes in $X(r)$ for the DM-influenced model, with a radiative core, are much smoother.\\

Therefore, a convective core leaves a signature in the radial profile of the mean molecular weight and in the pressure modes oscillation frequencies of a pulsating star (see~\cite{2007ApJ...666..413C,2011A&A...529A..10C}). This signature builds as evolution proceeds on the main sequence and is caused by the discontinuity in the composition and hence in the sound speed at the edge of the growing convective core. This effect is clearly illustrated in Figure~\ref{fig-multirad}.c), where the sound-speed profiles in the inner regions of two representative models with and without a convective core are shown. The model with a convective core presents a distinct sound-speed discontinuity at the border between the convective region and the radiative one.\\

Particular combination of oscillation frequencies, the so-called diagnostic tools, may be able to isolate the signature left on the frequencies by the presence of a small convective core. One such diagnostic tool is $dr_{0213}$, which combines frequencies with modes of degree $l = 0$ to 3 \cite{2007ApJ...666..413C,2014MNRAS.438.1751B}. 
Since $l = 3$ modes may not be always observable, specially from space-based data, other diagnostic tools are preferable, such as $r_{010}$~\cite{2003A&A...411..215R}:
\begin{equation}
r_{010} = \{r_{01}(n), r_{10}(n), r_{01}(n+1), r_{10}(n+1), ...\} \, ,
\end{equation}where:
\begin{equation}
r_{01}(n)={(\nu_{n-1,0}-4\nu_{n-1,1}+6\nu_{n,0}-4\nu_{n,1}+\nu_{n+1,0})\over 8\, (\nu_{n,1}-\nu_{n-1,1})} \, ,
\end{equation}
\begin{equation}
r_{10}(n)={-(\nu_{n-1,1}-4\nu_{n,0}+6\nu_{n,1}-4\nu_{n+1,0}+\nu_{n+1,1})\over 8\, (\nu_{n+1,0}-\nu_{n,0})}  \, .
\end{equation}
The frequency derivative of $r_{010}$ may be used to diagnose the presence or the absence of a convective core in the star and, when present, it may in principle provide information about convective core's properties~\cite{2014MNRAS.438.1751B}.\\ 
\begin{figure*}
\centering
\includegraphics[scale=0.6]{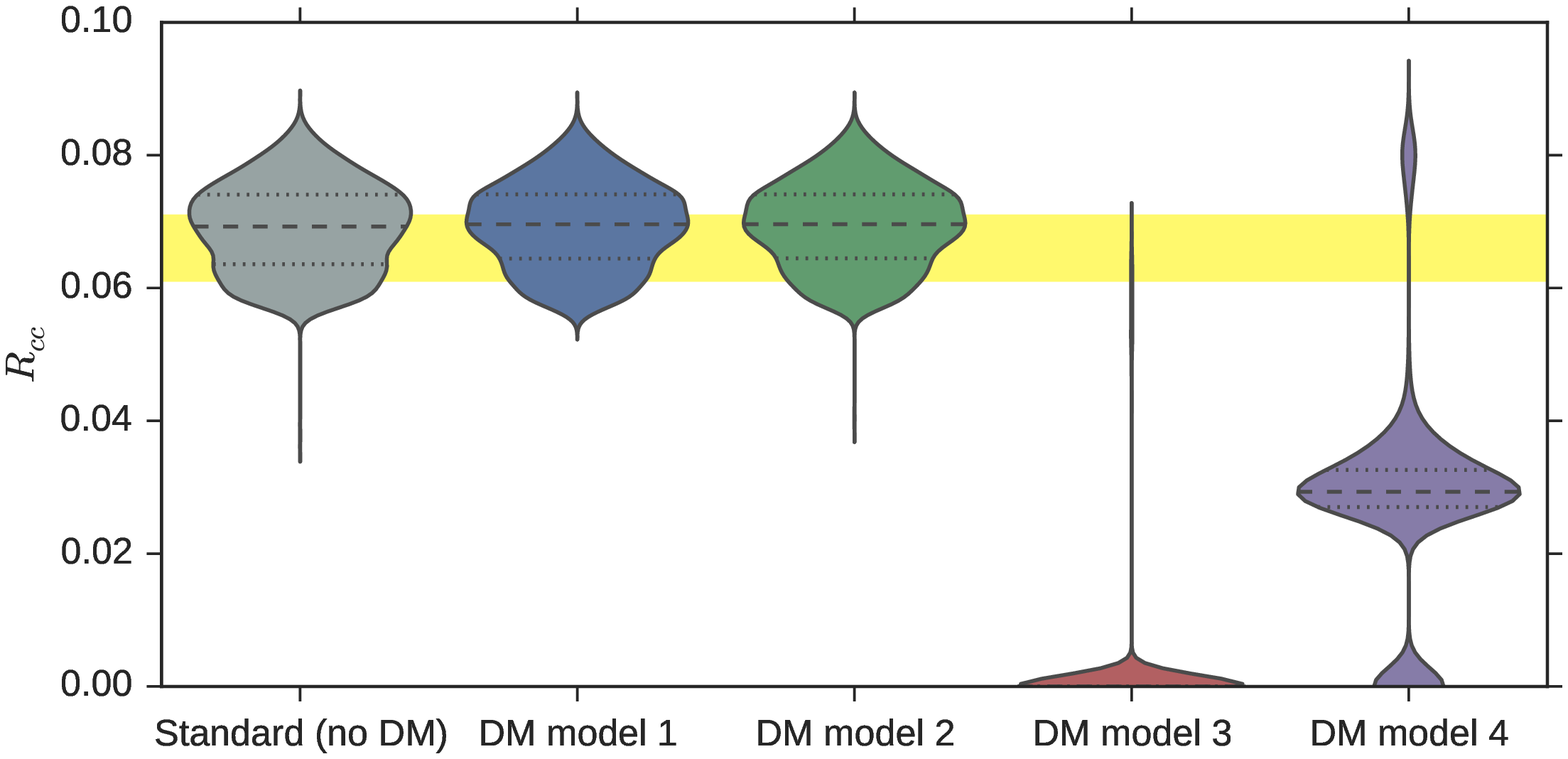}
\caption{Distribution of the radii of the convective core, $R_\mathrm{cc}$, in all the computed models that reproduce the observations of the star \textit{Dushera}, in the standard scenario without DM (grey) and including the impact of different DM models (colors, see text). The shapes of the distributions (violin plots) show the probability density, and the internal discontinuous lines correspond to the first quartile, the median and the third quartile. The yellow area shows the range of the $R_\mathrm{cc}$ inferred for \textit{Dushera} in Ref.~\cite{2013ApJ...769..141S}.}
\label{fig-rcc}
\end{figure*}

In this work we considered the slope of the seismic parameter $\langle\Delta\nu\rangle\,r_{010}$, where $\Delta \nu_{n,l} = \nu_{n+1,l} - \nu_{n,l}$ is the large frequency separation defined as the difference in frequency between modes of the same degree $l$, and consecutive radial order $n$. It was shown that this frequency derivative is sensitive to the frequency perturbation induced by the discontinuity in sound speed at the edge of a small convective core and can be used to infer the amplitude of the relative sound-speed variation at that region~\cite{2011A&A...529A..10C,2014MNRAS.438.1751B}.\\

The absolute value of the slope of the above-mentioned diagnostic tool, $\left| S\{\Delta\nu\,r_{010}\}\right|$, was computed for all models in our grids in the range of the observed frequencies~\cite{2013A&A...560A...2R} using the definitions from Ref.~\cite{2014MNRAS.438.1751B}. The results of our computations confirm that this slope is a good tracer of the presence/absence of a convective core in the stellar models.\\

\section{Analysis and results}
\label{sec-results}

Here we present the analysis of the properties of all the stellar models, computed and selected as described in Section~\ref{sec-modeling}, their oscillations, and their comparison with the observations of \textit{Dushera}. The reader can have a quick picture of our results by reading Table~\ref{tab-params}, where the average and standard deviations of the classical and seismic properties of the models are shown.\\

\begin{figure*}[!t]
\centering
\includegraphics[scale=0.6]{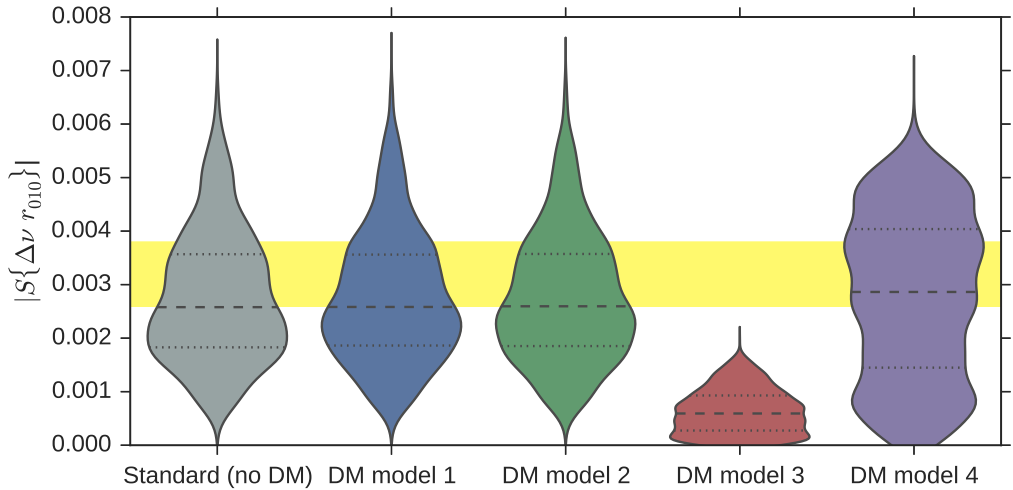}
\caption{Distribution of the asteroseismic parameter $|S\{ \Delta\nu \, r_{010}\}|$ in all the computed models that reproduce the observations of the star \textit{Dushera}, in the standard scenario without DM (grey) and including the impact of different DM models (colors, see text). The shapes of the distributions (violin plots) show the probability density, and the internal discontinuous lines correspond to the first quartile, the median and the third quartile. The yellow area shows the value of $|S\{ \Delta\nu \, r_{010}\}|$ calculated from the observed frequencies with its uncertainty.}
\label{fig-slLsr010}
\end{figure*}

In addition, the distribution of the radius of the convective core $R_\mathrm{cc}$ and the seismic parameter $|S\{ \Delta\nu \, r_{010}\}|$ obtained in all the models (DM-free and DM-influenced) can be easily compared with the $R_\mathrm{cc}$ derived in Ref.~\cite{2013ApJ...769..141S} and the observed $|S\{ \Delta\nu \, r_{010}\}|$ in Figures~\ref{fig-rcc} and~\ref{fig-slLsr010}, respectively. These figures depict \textit{violin plots}, a convenient method of visually showing the distribution of data. The width of the violins is proportional to the number of models obtained with a given $R_\mathrm{cc}$ or $|S\{ \Delta\nu \, r_{010}\}|$, whereas the internal discontinuous lines show the first quartile, the median and the third quartile.\\

For each class of DM model, the $\chi^2$'s of the properties of the stellar models were calculated:
\begin{equation}
 \chi^2_Q = \left( \frac{Q_{obs} - \bar{Q}_{mod}}{\sigma_{Q_{obs}}} \right)^2 \;,
\end{equation}for $Q=\{T_\mathrm{eff}, \log (g), [Z/X]_s, |S\{ \Delta\nu \, r_{010}\}|\}$. The individual $\chi^2_Q$, the combined $\chi^2_T = \sum_Q \chi^2_Q$, and the $p$ values computed assuming Gaussian errors, are shown in Table~\ref{tab-chi2p}.\\

\begin{table}[!t]
\centering
 \begin{tabular}{l r r r r r r}
 \hline\hline
  & $\chi^2_{T_\mathrm{eff}}$ & $\chi^2_{\log (g)}$ & $\chi^2_{[Z/X]_s}$ &  $\chi^2_{|S\{ \Delta\nu \, r_{010}\}|}$ & $\chi^2_T$ & $p$ \\
 \hline
 Standard & 0.00 & 0.17 & 0.00 & 0.51 & 0.68 & 0.88 \\
 DM model 1 & 0.00 & 0.17 & 0.00 & 0.53 & 0.70 & 0.87 \\
 DM model 2 & 0.00 & 0.17 & 0.00 & 0.52 & 0.69 & 0.88 \\
 DM model 3 & 0.07 & 0.17 & 0.03 & 18.35 & 18.62 & 0.0003 \\
 DM model 4 & 0.01 & 0.17 & 0.01 & 0.47 & 0.66 & 0.88 \\ 
 \hline\hline
 \end{tabular}
  \caption{$\chi^2$'s and $p$ values of the models of $Dushera$.\label{tab-chi2p}}
\end{table}

First, we found that the modeling of \textit{Dushera} without DM reproduced with an excellent agreement the results of previous analysis on the same star~\cite{2013ApJ...769..141S}, confirming the robustness of our grid and selection procedures. All the DM-free models of \textit{Dushera} were found to have a small convective core, with a mean radius of 0.069~R$_{\star}$ and a standard deviation of 0.006~R$_{\star}$. The seismic parameter $|S\{ \Delta\nu \, r_{010}\}|$, used here as a diagnostic tool of the stellar interior, was also found to be in agreement with the value calculated from the observed oscillation frequencies.\\

When the influence of DM particles with the properties of DM models 1 and 2 were considered, we obtained essentially the same results as in the standard scenario, \textit{i.e} the model without DM (see Tables~\ref{tab-params} and~\ref{tab-chi2p}). This result implies that these models of DM particles do not significantly impact the internal structure of the star. Accordingly, the distributions of the obtained radii of the convective cores found in all the models, as well as their seismic properties, are very similar to those obtained in the DM-free scenario, as shown in the violin plots of Figures~\ref{fig-rcc} and~\ref{fig-slLsr010}.\\

On the other hand, when the influence of the existence of DM particles with the properties of model 3 was taken into account, we found that only 2\% of the models of \textit{Dushera} had a convective core. This is a consequence of the efficient energy transport by DM conduction, which reduces the temperature gradient in the core of the star and prevents convection.\\

As expected, this significant modification in the structure of the stellar core had a clear impact in the stellar oscillations. The diagnostic tool $|S\{ \Delta\nu \, r_{010}\}|$ was found to be very sensitive to the absence of a convective core in the stellar models. As shown in Tables~\ref{tab-params}, \ref{tab-chi2p} and in Figure \ref{fig-slLsr010}, the parameter $|S\{ \Delta\nu \: r_{010}\}|$ for the DM model 3 is clearly in disagreement with the observations. Whereas the DM-free and DM models 1 and 2 reproduce $|S\{ \Delta\nu \: r_{010}\}|$ within one standard deviation, the mean $|S\{ \Delta\nu \: r_{010}\}|$ for DM model 3 shows a discrepancy higher than 4-$\sigma$ ($\chi^2_{|S\{ \Delta\nu \, r_{010}\}|}>18$). Assuming Gaussian errors, the $p$ values from $\chi^2_T$ are $p=0.0003$ for this model of DM and $p\simeq0.88$ for the other models, indicating a good fit to the data for all the models except the DM model 3, which can be excluded with a 99\% confidence level.\\

In the case of DM model 4, the WIMP-proton scattering cross-section is so large that the DM mean free path inside the star becomes very small and DM is in local thermal equilibrium with the stellar plasma. In this scenario the impact of DM conduction is drastically reduced, being noticed only by a small contraction on average of the convective core size when compared with the standard scenario. This impact is not sufficient to leave a clear signature in the diagnostic parameter $|S\{ \Delta\nu \, r_{010}\}|$ and does not lead to a goodness of fit significantly different from the DM-free scenario.\\

\section{Discussion and conclusions}
\label{sec-concl}
Our results show that the existence of asymmetric DM particles with the properties that explain the positive results in the CoGeNT experiment would lead to the suppression of the convective core recently detected in the main-sequence star \textit{Dushera}. We have demonstrated that the stellar oscillations, in particular the diagnostic tool $|S\{ \Delta\nu \, r_{010}\}|$, are sensitive to the stellar core and can be used to exclude with a 99\% confidence level the existence of this particular model of DM particles. Whereas the observed non-seismic properties of \textit{Dushera} ($T_\mathrm{eff}$, $\log (g)$, $[Z/X]_s$) can be accurately reproduced by the DM-impacted stellar models, the parameter $|S\{ \Delta\nu \, r_{010}\}|$ shows a mean discrepancy with the measured stellar oscillations above 4-$\sigma$.\\

The other models of ADM particles studied here, namely those that would explain the signals in DAMA and CDMS-Si experiments, were found not to lead to significant modifications in the stellar structure, and consequently their existence cannot be disproved using our approach. Even though we used the positive signals in direct detection experiments to set our benchmark models, we note that our approach could be applied to any model of DM particles with similar properties.\\

In our analysis we included the systematics that arise due to the variations of the major standard mechanisms that are known to impact the presence or absence of a convective core in a star, namely its mass, age, metallicity, initial helium abundance, and the parameters that control the efficiency of overshooting and convective mixing. In addition, to grasp the impact of changes in the stellar composition, we re-computed our grid in the DM-free scenario using the old estimation of the solar chemical mixture~\cite{1998SSRv...85..161G}, resulting in a variation of the mean $R_\mathrm{cc}$ of only 10$^{-3}\;$R$_{\star}$. These results confirm the robustness of our approach. Nonetheless, we note that unexplored input physics in the modelling may introduce further systematic errors (see \textit{e.g.} Ref~\cite{2014A&A...569A..21L}). Regarding the modelling of the DM impact, the main additional uncertainty comes from the DM density around the star, which we assumed equal to the local solar value. The uncertainty in this quantity (see \textit{e.g.} Refs.~\cite{2014JPhG...41f3101R,PatoIoccoBertone2015}) directly influences the DM-proton scattering cross-sections assumed for the studied DM models.\\

The method proposed here provides a novel strategy to test the existence of specific models of yet undiscovered particles, candidates to constitute the DM of our Universe. Although the sensitivity of this method to the DM-nucleon scattering cross sections is below that of next-generation direct detection experiments, it constitutes a valuable complementary approach to probe the low-mass region of the WIMP DM parameter space.\\ 

\acknowledgements
J.C. acknowledges the support from the Alexander von Humboldt Foundation. I.M.B. acknowledges the support of the Funda\c c\~ao para a Ci\^encia e a Tecnologia (FCT) in the form of the grant reference SFRH/BPD/87857/2012 and acknowledges the support from the EC Project SPACEINN (FP7-SPACE-2012-312844).

%

\end{document}